\documentclass[12pt]{article}

\usepackage{mathrsfs}
\usepackage{amsmath}

\global\arraycolsep=1pt \oddsidemargin .20in \evensidemargin .5in
\usepackage[Symbol]{upgreek}
\usepackage{bbm}
\usepackage{dsfont}
\usepackage{amssymb}
\usepackage{textcomp}
\newcommand{\ba}{/ \hspace{-1.2ex}}

\newcommand{\CR}{\nonumber \\*}

\newcommand{\trace}{\hbox {Tr}~}

\DeclareMathAlphabet{\mathpzc}{OT1}{pzc}{m}{it}

\def\d{\mathfrak{d}}

\usepackage[colorlinks=true,backref=true,linkcolor=black,anchorcolor=black,citecolor=black,filecolor=black,menucolor=black,pagecolor=black,urlcolor=black]{hyperref}

\usepackage[Symbol]{upgreek}
\usepackage{caption}
\usepackage{bbm}
\usepackage{dsfont}
\usepackage{amssymb}
\usepackage{textcomp}

\def\d{\delta}

\def\be{\begin{equation}}
\def\ee{\end{equation}}
\def\bea{\begin{eqnarray}}
\def\eea{\end{eqnarray}}
\def\bdis{\begin{displaymath}}
\def\edis{\end{displaymath}}
\def\corr{$\clubsuit$}
\def\nn{\nonumber}

\def\d{\delta}

\def\A{\mathscr{A}}

\begin{document}

\newcommand{\dslash}{\partial\!\!\!/}
\newcommand{\aslash}{A\!\!\!/}
\newcommand{\Dslash}{D\!\!\!\!/}
\newcommand{\pslash}{p \hspace{-1.7mm} /}
\newcommand{\kslash}{k \hspace{-1.7mm} /}
\newcommand{\bs}{b \hspace{-1.7mm} /}

\allowdisplaybreaks[1]
\renewcommand{\thefootnote}{\fnsymbol{footnote}}
\def\corr{$\spadesuit $}
\def\trefle{ $\clubsuit$}

\renewcommand{\thefootnote}{\arabic{footnote}}
\setcounter{footnote}{0}


 \def\stop{$\blacksquare$}
\begin{titlepage}
\null
\begin{flushright}
CERN-PH-TH/2010 
\end{flushright}
\begin{center}
{{\Large \bf
 SU(5)-invariant 
   decomposition 
of ten-dimensional    Yang--Mills supersymmetry
 }}
 \lineskip .75em \vskip 3em \normalsize {\large Laurent
 Baulieu\footnote{email address: baulieu@lpthe.jussieu.fr}
\vskip 1em
{\it Theoretical Division CERN}\footnote{
CH-1211 Gen\`eve, 23, Switzerland }
\\
{\it LPTHE  Universit\'e Pierre et Marie Curie
}\footnote{ 4 place Jussieu, F-75252 Paris
Cedex 05, France.}
 }
%








\end{center}
\vskip 1 em
\begin{abstract}

The N=1, d=10 superYang--Mills action is   constructed in a twisted form, using 
$SU(5)$ invariant decomposition of spinors in 10 dimensions. The action 
and its  off-shell closed twisted scalar supersymmetry operator $Q$
derive from a Chern--Simons term.  The action  can be decomposed as       the sum of a          term  in the 
cohomology  of $Q$ and  of a term that is  $Q$-exact.  The first term is a fermionic Chern-Simons term for a twisted component of the   Majorana--Weyl gluino and  it  is  related to   the second one   by a twisted vector supersymmetry with 5 parameters.
     The cohomology of $Q$ and some topological observables are defined  from descent equations. 
In this $SU(5)\subset SO(10)$ invariant decomposition,   the N=1, d=10 theory is determined by   
   only  6 supersymmetry generators, as in  the twisted N=4, d=4 theory. There is  a   superspace with 6 twisted fermionic directions,  with solvable constraints.  
 \end{abstract}

\end{titlepage}
 \def\Tr{\trace}
 
 \section{Introduction}

 There is a huge literature concerning the construction of supersymmetric theories in a twisted form.    In twisted formulations    one  often gets  great simplifications of supersymmetric transformations, and    sets of auxiliary fields can be found for  smaller  supersymmetric subalgebra, which are however big enough to determine the theory. The twisted formulations are also the privileged framework to define topological observables. 
 
 The case of maximal supersymmetry has been studied quite extensively, in 4, 6 and 8 dimensions, but   the   twist of    the N=1, d=10  theory has not been studied in depth.  
 In fact,   twisted aspects of theories with N=1 supersymmetry are less familiar than those of 
 theories with extended symmetries. The latter theories  seem  easier to twist,  using  the possible  mixings between the  R-symmetry  and the Lorentz symmetry.  

   For   N=1 theories, there is no  $R$ symmetry to  be used. However,  in  K\"ahler  manifolds,   the Lorentz symmetry is   reduced down to $SU(D)\subset SO(2D)$, so that    the    spinors can be decomposed into  holomorphic and antiholomorphic forms and a twist can be often  performed.  We refer to 
 \cite{Johansen} 
\cite{Witten1}      
\cite{popov}
\cite{Hofman}
\cite{bata}   \cite{recons}\cite{N4twisted}    \cite{bana}  for works related to  N=1 theories in twisted form, and to \cite{10DSYM} for a  $Spin(7)\subset SO(10)$ invariant off-shell  description in d=10. 
     
     Here we      directly   build 
the N=1, d=10 superYang--Mills  theory  in a twisted form, using  
$SU(5)$ invariant decomposition of Majorana--Weyl spinors in 10 dimensions. We    build the action and its symmetries by using   a   unification between   twisted components of  the  Majorana--Weyl gluino  and  the antiholomorphic part of    the gauge field.  In this construction, the Batalin--Vilkowiski formalism is of great help. A  Chern--Simons action  with  off-shell closed twisted scalar supersymmetry operator $Q$   emerges quite naturally,  using   a generalization of previous works \cite{bv}.   A striking property is found :  the  N=1, d=10 action   is the sum of a term in the 
cohomology  of $Q$ and  of  a $Q$-exact term.    Both terms are related by a twisted vector supersymmetry.  Off-shell closure is obtained modulo Yang--Mills  gauge invariance, but  can  be enforced exactly by introducing shadows \cite{bana}.  

The twisted scalar and vector supersymmetries   that we exhibit can be also  obtained  
  by a  brute force twist of   the  known on-shell closed 10-dimensional superYang--Mills supersymmetry transformations, using     $SU(5)\subset SO(10)$ decompositions of    fields and symmetry generators.  In this way one misses however the        geometrical aspects.   The main point of this paper  is in fact the central role played by the Chern--Simons term and the   observation that the N=1, d=10 superYang--Mills  action is determined by only 6 twisted generators, with a possible   twisted superspace 6=1+5  fermionic directions. The later  can be constructed as in      \cite{bana}\cite{10DSYM}\cite{ twistsp}.

     \section{ The  d=10 superYang--Mills action as a supersymmetric gauge-fixing of the fermionic  Chern--Simons  action 
 $\chi_{0,2} D_{0,1} \chi_{0,2}$}

 \def\T {{\rm Tr\ }}
 
 \def\bm{{\bar m}}
  \def\bn{{\bar n}}
   \def\bp{{\bar p}}
    \def\bq{{\bar q}}   
  \def\br{{\bar r}}
    \def\bs{{\bar s}}
\font\mybb=msbm10 at 12pt
\font\mybbb=msbm10 at 8pt
\def\bb#1{\hbox{\mybb#1}}
\def\bbb#1{\hbox{\mybbb#1}}
\def\complex{{\bb{C}}}
\def\ccomplex{{\bbb{C}}}

\def\A{{  \cal A}}
\def\D{{  \overline D   }}

 \def\ba{{\bar a}}
  \def\bbar{{\bar{ b}}}
   \def\bc{{\bar c}}
    \def\bd{{\bar d}}   
  \def\bee{{\bar e}}
    \def\bf{{\bar f}}

The 16 components of  10-dimensional Majorana--Weyl spinor can be decomposed in a  $SU(5)\subset SO(10)$ invariant way as a   set of holomorphic and antiholomorphic forms 
\bea
\lambda \sim  \chi_{0,0}\equiv\chi,  \Psi _{1,0}\equiv dz^m \Psi_ {m}, \chi_{0,2}\equiv \chi_{\bm\bn}dz^\bm dz^\bn 
 \eea
 with $16=1\oplus5\oplus 10$.  The $z^m$ are complex coordinates, with   $SU(5)$  indices $m,n,p,..$   running from 1 to 5. 
 $z^\bm,z^\bn,..$ are their complex conjugates. $\Omega_{5,0}$  is      is the  complex structure of the manifold, 
 $d^{10} x= \Omega_{5,0}  dz^{\bar 1}dz^{\bar 2}dz^{\bar 3}dz^{\bar 4}dz^{\bar 5}$. The K\"ahler 2-form 
 is   $J_{\bn m }= - J_{m\bn} $,  $J_{mn }= J_{\bm\bn} =0$, and  $J\equiv J_{m\bn }dz^m d\bar z ^\bn$. It can be used a  metrics, with the notation 
  $X_mY_{\bm}\equiv g^{m \bn  } X_mY_{\bn}$.
 
The 10-dimensional Dirac Lagrangian   can be   written in a twisted form   as 
 \bea\label{dirac}
 \Tr 
\lambda    \Dslash     \lambda =\Tr (\epsilon_{mnpqr}
\chi_{\bm\bn} D_\bp \chi_{\bq\br} 
 +  \chi_{\bm\bn} D_m \Psi_n  +\chi D_\bm \Psi _m)
 \eea
  Moreover,   the Yang--Mills Lagrangian
  $\Tr F_{\mu\nu} F ^{\mu\nu}$  can be written as 
 \bea\label{ym}
  \Tr(
    F_{mn} F _{\bm\bn} +  (F^m_\bm)^2
  \sim
    \Tr( F_{mn} F _{\bm\bn} -\frac { h^2}{2} +h F^m_\bm)
 \eea
  modulo a boundary term  
  $J{ \small \wedge}J{ \small \wedge}J{ \small \wedge}\Tr (F{ \small \wedge}F)$. 
 Here  $h$ is an auxiliary field and      $A= A_{1,0} +A_{0,1}$,  $A_{1,0}=  A_m dz^m$  $ A_{0,1}=A_\bm dz^\bm$.
 So the d=10 supersymmetric Yang--Mills action can be written  in a twisted form as the sum of  both expressions~(\ref{dirac}) and  ~(\ref{ym}).     
  The aim is to directly build this sum and its symmetries in a TQFT formalism.
  
 The action will be expressed as   the sum of a  term in the cohomology of $Q$ and    of  a  $Q$-exact term, where $Q$ is a scalar twisted supersymmetry generator. Moreover, the former is the 10-dimensional projection  of a  fermionic Chern--Simons term. The occurrence of  a term in  the cohomology of $Q$ is specific to   10-dimensions. Dimensionally reduced actions with maximal supersymmetry, such as the d=4, N=4 theory, are purely   $Q$-exact terms  in lower dimensions.
 
We will  unify the fields $A_\bm $ and  $\chi_{\bm\bn}$ as elements of a generalised 1-form   
 $\cal A $ and  find   a closed nilpotent scalar supersymmetry that acts naturally on this 1-form.  This   field   unification within   Chern--Simons or  BF  theories   has been noticed in other papers \cite{bv}.
 The method  will determine at once      the supersymmetry  and the twisted action.

To proceed, we    consider the  following unified one-form that is made of purely antiholomorphic forms, all valued in the Lie algebra of a given gauge group. 
\bea{  \cal A} =^*c_{0,5} ^{-4} +
^*A_{0,4 }^{-3} +
^* \chi_{0,3} ^{-2}  
 +\chi_{0,2}   ^{-1} 
+A^{0}_{0,1}  +c^1
\eea
The grading is sum  of 
 the above index  that  is the shadow number (it was    called ghost number as in the old TQFT language) plus the ordinary form degree.  The notation  $^*\varphi $  means that $^*\varphi $ is a Batalin--Vilkowiski  (BV)  antifield  \footnote
 {
 For the 3-dimensional Chern--Simons action,  the last reference of \cite{bv}
   defined   ${\cal A} =^*c_{ 3} ^{-2} +
^*A_{ 2 }^{-1} + 
+A^{0}_{ 1}  +c^1$, where  $^*A_{ 2 }^{-1}$ and  $^*c_{ 3} ^{-2}$  are  the antifield of the gauge field $A$ of the Faddeev Popov ghost $c$.
One can presumably generalize the construction   with $SU(N)$,  $N>5$.}.  If one uses indices, one has
$
{  \cal A} = 
^*c^{}_{\bm\bn\bp\br\bs}dz^\bm dz^\bn dz^\bp dz^\br dz^\bs  +
^*A_{\bm\bn\bp\br }dz^\bm dz^\bn dz^\bp dz^\br+
{^*\chi}_{\bm\bn\bp }dz^\bm dz^\bn dz^\bp 
 +\chi_{\bm\bn}^{-1}  dz^\bm dz^\bn
+A_\bm dz^\bm +c^{1} 
$.
We then consider  the Chern--Simons form
\be
\Delta=  \Tr (\A  d\A  +\frac{2}{3} \A\A\A)
\ee
and its action projected on a  d=10 manifold with  holonomy $SU(5)\subset SO(10)$,   
\bea
I= \int   \Omega _{5,0}  \Delta
=\int   \Omega _{5,0}\ \Tr  \big (
\chi_{0,2} \D \chi_{0,2}
+  {^*\chi}_{0,3 }(F _{0,2} +[c,  \chi_{0,2}])
-  {^*A}_{0,4 }\D c
- {^*c}_{0,5 }cc
\big )
\eea
 where $\D\equiv dz^\bm d_{A_\bm}= dz^\bm (  \partial_\bm +{A_\bm} )$. Because of the Chern--Simons formula $d \Delta = \Tr  F_\A  F_\A$, $I$  satisfies a master equation, and can be  interpreted as a BV action.
   The nilpotent symmetry of the action, expressed by the fermionic generator $\delta$,  is obtained in a standard way by the       generalized  equations of motion of $I$, 
\bea\label{BV}
\delta \varphi= \frac{\delta I}{\delta ^*   \varphi}\ \ \ \  \delta ^*\varphi= -\frac{\delta I}{ \delta  \varphi}\eea
$I$  needs  a     BV gauge-fixing, that is, the introduction of a gauge-fixing function  $\cal Z[ \varphi  ]  $ for fixing 
 the  antifields~$^*\varphi$ in function of the    $ \varphi$'s and  get a quantum field theory  for   $ \varphi$
\bea\label{BVs}
 ^*\varphi= \frac{\delta \cal Z[ \varphi  ]}{\delta  \varphi}
 \eea
  The   choice of the local  functional
 $\cal Z[ \varphi  ]$   will  be justified by power counting and symmetry requirements, with  the demand  of a vector symmetry $Q_\bm$  of the gauge-fixed action that anticommutes with $Q$. This will warrantee  a    Poincar\'e  supersymmetry interpretation.

The  antifield independent part     $   \Omega _{5,0}\Tr \chi_{0,2} \D \chi_{0,2}$  of $I$ can be called the   ``classical"  Lagrangian.
      It is        a fermionic generalisation of the Chern--Simons action.  It   is $Q$-supersymmetric,  with   $Q\chi_{0,2} =F_{0,2}$, because of $ \D F_{0,2}=0$. 
      It can be completed by addition of the topological term 
     $JJJ\Tr FF$, so that,    we understand the relevance of the ``topological"  Lagrangian 
 \bea   \Omega _{5,0}\Tr \chi_{0,2} \D \chi_{0,2}
 +     JJJ\Tr FF  \eea
Had we started from such a topological Lagrangian and not understood the Chern--Simons structure, more work would have been needed to
understand conventionally its TQFT gauge-fixing   into   the   N=1, d=10 theory. 

In the BV formalism, the so-called antifields ${^*\chi}_{0,3 },    {^*A}_{0,4 } $ and 
$ ^* c_{0,5 }$  are,   respectively,   the sources of    $\delta$-supersymmetry transformations  of the fields $\chi_{0,2}$,  $A_{0,1} $ and  $c$. 

Eqs~(\ref{BV}) give  the scalar  supersymmetry of  the action $I$, with    
\be   \begin{split}
\delta A_\bm  &= D_\bm c ,       \ \ \ \ \ \ \ \    \ \ \ \ \ \ \ \    \delta c   = -cc  \\
\delta \chi_{\bm\bn} &= F_{\bm\bn}  -  [c, \chi_{\bm\bn}]  \\
\end{split}\hspace{10mm}
\ee
\be   \begin{split}
\delta{^*\chi}_{0,3 }    &= \D  \chi_{0,2}  -[c,   {^*\chi}_{0,3 }]     \\
\delta {^*A}_{0,4 }   
  &=   \D  {^*\chi}_{0,3 }  -       [\chi_{0,2}  ,   \chi_{0,2}]  -  [c,    {^*A}_{0,4 } ]      \\
\delta  {^*c}_{0,5 } &=   \D   {^*A}_{0,4 } -  [\chi_{0,2},   {^*\chi}_{0,3 }  ] -    [c,   {^*c}_{0,5 } ]       \\ 
\end{split}\hspace{10mm}
\ee
 
The property  $\delta^2=0$ is ensured  by construction. The equivariant operator  $Q$ is obtained by molding out    the Yang--Mills symmetry,  that is,   by setting    $c= {^*c}_{0,5 }=0$.   With  this simplification, the  BV   Chern--Simons action is   \bea  I=\int   \Omega _{5,0} \ \Tr   (
 \chi_{0,2} \D \chi_{0,2}
+  {^*\chi}_{0,3 }F _{0,2} 
)
\eea 
Its nilpotent  twisted scalar supersymmetry  generator $Q$ is
 \be   \begin{split} 
Q A_{0,1}   &= 0    \ \ \        \ \ \    \ \ \ \      \ \ \ \ \ \         Q {^*A}_{0,4 }   
  =   \D  {^*\chi}_{0,3 }  -       [\chi_{0,2}  ,   \chi_{0,2}]              \\
Q\chi_{0,2} &= F_{0,2} \ \ \ \ \ \    \ \ \    \ \ \ \  Q{^*\chi}_{0,3 }    = \D  \chi_{0,2}                                  \
\end{split}\hspace{10mm}
\ee
 The antifield $* B_{0,3 }$  has a  2-form gauge invariance, due to the Bianchi identity. 
 \be   \begin{split}
 s\chi_{0,2} &= \D L _{0,1}   \\
s {^*\chi}_{0,3 }    &= \D  M_{0,2}     +  [  L _{0,1},  \chi_{0,2}]     \\
 \end{split}\hspace{10mm}
\ee
 with $sI=0$. It  has a ghost of ghost of ghost   degeneracy 
 $ M_{0,2} \sim M _{0,2}+ \D M_{0,1},  M_{0,1} \sim M _{0,0}+ \D M_{0,0} $.
This explains   that the antifield   ${^*\chi}_{0,3 }$  truly  counts for    4 =10-5+1 degrees freedom. After BV gauge-fixing,  ${^*\chi}_{0,3 }$ will  be expressed in function of an a holomorphic 1-form $A_m$, which  also counts for  
 4=5-1  degrees freedom, modulo  the gauge-invariance $A_m\sim A_m +D_m \epsilon$.  
 This gauge symmetry will be  preserved by    the BV gauge function~$\chi_{0,2} F_{2,0}$.

 
To eliminate the antifields, and obtain a theory with  well-defined propagators for the fields (modulo the ordinary Yang--Mills  gauge invariance),  the standard BV routine suggest one to  introduce two trivial BRST doublets, $A_m, \Psi_m$, and $\chi,h$. Their antifields are 
$ ^*A_\bm, ^*\Psi_\bm $, and $ ^*\chi,^*h $. One  thus adds    the following  trivial action $I'$ to  $I$ 
\bea
I'=  \int  \Tr (    ^*A_\bm  ( \Psi_m -D_m c)  +^*\chi (h-   [c,\chi ])  -   ^*h[c,h])
)
\eea
 Eqs.~(\ref{BV})  give then the $\delta$-transformations  for these doublets        (and    simple modifications for   $  \delta  {^*c}_{0,5 }$ because of the $c$ dependance in $I'$)
\be   \begin{split}
\delta A_m  &= \Psi_m - D_m c      \ \ \ \ \ \    \delta^* A_\bm    =    -[c,   ^* A_\bm]       \\
\delta \Psi _m  &= -[c,\Psi_m ]       \    \ \ \ \ \ \     \ \    \delta^* \Psi_\bm     =   ^* A_\bm     -[c,   ^* \Psi_\bm ]                      \\
\delta \chi  &= h   -  [c, \chi ]      \    \ \ \ \ \ \    \ \ \  \        \delta   ^*\chi  =   -  [c, ^*\chi]       \\
\delta h   &= -[c, h]                  \  \  \ \ \ \ \ \    \ \ \  \  \  \      \delta   ^*h =   -  [c, ^*h]   +  ^* \chi           \\   
\end{split}\hspace{10mm}
\ee
For    $c=0$, one gets the Q supersymmetry 
  \be   \begin{split}
Q A_m  &= \Psi_m   \ \ \ \ \ \  Q\Psi _m   =0 \\
Q\Psi _m  &=0      \ \ \ \ \ \     \ \      Q^* \Psi_\bm     =   ^* A_\bm \\
Q \chi  &= h      \ \ \ \ \ \    \ \ \  \   Q  ^*\chi  =    0      \\
Q h   &=0      \ \ \ \ \ \        \ \ \  \    Q   ^*h =    ^* \chi   \\
\end{split}\hspace{10mm} \ee
Finally, 
 $Q$  can  be derived from  the     equivariant action
\bea
I_T=  \int \Tr(     \Omega _{5,0} \ \Tr   (
 \chi_{0,2} \D \chi_{0,2}
+  {^*\chi}_{0,3 }F _{0,2} )   +\Tr\  (
   ^*A_\bm   \Psi_m    +^*\chi h   
)
\eea

 We  choose the  following  $s$-invariant  BV gauge-fixing fermion
\bea
{\cal Z_V }=   \T ( \chi_{\bm\bn}   F_{mn}  +  \chi (\frac{h}{2}  +  F_{m\bm})  ) 
\eea
Then, Eq.~(\ref{BVs})   fixes the antifields in functions of the fields
 \be   \begin{split}\label{anti}
^*B_{\bm\bn\bp}  & = \epsilon _{\bm\bn\bp\br\bs}   F_{rs}  \\
^* A_\bm   &=    D_n  \chi _{\bn \bm}+    D_\bm  \chi\\
^*\chi  &= \frac{h}{2}  +  F_{m\bm}    \\ 
^*h &=0\\
\end{split}\hspace{10mm}
\ee
We will              see shortly that this gauge-fixing function implies an additional  $SU(5) $-vector symmetry with 5 parameters for the complete gauge-fixed action, which gives a larger symmetry with 6=1+5 parameters.

By substitution of   the antifield values~(\ref{anti}) in $I_T$,    one finds  the      twisted N=1, d=10 action, modulo the topological term $\int JJJ  \Tr (FF)$
 \bea\label {twisted}
I_T=  \int d^5zd^5\bar z\T (    F_{mn} F _{\bm\bn} -\frac { h^2}{2} +h F^m_\bm+\epsilon_{mnpqr}
\chi_{\bm\bn} D_\bp \chi_{\bq\br} 
 +  \chi_{\bm\bn} D_m \Psi_n  +\chi D_\bm \Psi m
)\nn
\CR
= \int  \Tr (  F_{\mu\nu} F ^{\mu\nu}+  
\lambda  \Dslash  \lambda)   -\int  JJJ \Tr (FF) )\ \ \ \ \ \ \ \ \ \ \ \ \  \ \ \ \ \ \ \ \ \ \  \ 
\eea
 
Moreover,  after this elimination of the antifields,   $I_T$   splits into two distinguished   terms
  \bea\label{I_V}
I_T=  \int   \T\big    (    
\epsilon_{mnpqr}
\chi_{\bm\bn} D_\bp \chi_{\bq\br}   +Q  ( \  
   \chi_{\bm\bn} F_{mn}   +\chi (\frac { h}{2} + F_{m\bm}) \ )\big 
)
\eea

The first Chern--Simons-like   term $\T \epsilon_{mnpqr}
\chi_{\bm\bn} D_\bp \chi_{\bq\br} $  is in the cohomology of $Q$ and the second term is Q-exact.
This decomposition of the N=1,d=10 Lagrangian  according to  the cohomology of $Q$
 is    quite interesting, and, moreover,  it has been derived from   the simplest
BV action  $\int  \Omega_{5,0}  \Tr   (
\chi_{0,2} D \chi_{0,2}
+  {^*\chi}_{0,3 }F _{0,2}) $.  Scalar  
  supersymmetry   is a genuine consequence of  the Bianchi identity, in a typically TQFT way.

 One has 10=9 (for $A$) +1(for $h$) bosonic  degrees of freedom,     modulo  the gauge invariance of  $A$.  This  equates    the number of fermionic degrees of freedom in $ \chi_{\bm\bn}, \Psi_m, \chi$,  which is also 10=4 (for $\chi_{\bm\bn} $)  +5 (for $\Psi _{m} $)+1(for $ \chi $),  if one counts 
4 degrees of freedom for 
  $ \chi_{\bm\bn} $,     taking  into account the gauge symmetry  $\chi_{\bm\bn}\sim  \chi_{\bm\bn}+ \epsilon _{\bm\bn\bp\bq\br} D_p M_{qr}$. 
%
    In a sense, the   $Q$-exact part of the  action    corresponds    to a balanced TQFT, with a gauge symmetry for $
\chi_{\bm\bn}$.  

Because of the obvious $U(1)$ shadow number symmetry (the shadow charges are 1 for    $\Psi$,  -1 for 
$\chi_{\bm\bn}$ and $\chi$,  and 0 for    all other fields),   one has  a global 
$U(5)=SU(5)\times U(1)$ symmetry that commutes with the $Q$ symmetry. However, the classical term $\chi\D\chi$ has shadow number 2, and violates this $ U(1)$ symmetry in   the Lagrangian. This may support   the attracting idea of  understanding the term  $\exp\int \chi_{0,2} \D \chi_{0,2}$ as an observable that can be  inserted  in the path integral\footnote{This methodology has been already applied in lower dimensions\cite{nikita}.} and also serves as a gauge-fixing for the gauge  symmetry of the $Q$-exact term. Since the shadow number is only a $SU(5)\subset SO(10)$-invariant concept, one can however adopt the pragmatic attitude that it has to conserved only modulo 2. 

%

%

 \def\bm{{\bar m}}
  \def\bn{{\bar n}}
   \def\bp{{\bar p}}
    \def\bq{{\bar q}}   
  \def\br{{\bar r}}
    \def\bs{{\bar s
    
    }}
\font\mybb=msbm10 at 12pt
\font\mybbb=msbm10 at 8pt
\def\bb#1{\hbox{\mybb#1}}
\def\bbb#1{\hbox{\mybbb#1}}
\def\complex{{\bb{C}}}
\def\ccomplex{{\bbb{C}}}

  \section{Vector supersymmetry}
   One has the following vector symmetry
  \be  
\begin{split}
 \delta_\bp A_m &= g_{\bp m }   \chi \ \  \    \ \ \ \ \ \  \ \ \  \ \ \ \ \ \ \ \ \ \ \  \delta_\bp  \chi =0     \CR
\delta_\bp A_\bm   &= \chi_{\bp \bm }\   \ \  \  \  \    \ \ \ \ \ \ \ \ \ \    \ \ \ \ \ \ \ \ \          \delta_\bp  h  = D_\bp \chi  \CR  
\delta_\bp \Psi _m &= F_{\bp m}  - g_{\bp m }  h     \    \ \ \ \ \ \ \ \ \ \    \delta_\bp\chi_{\bm\bn} = \epsilon_{\bm\bn\bp\bq\br} F_{qr}
\end{split}
\ee
It satisfies $\{ \delta_\bq, \delta_\bp\}=0 $ but   $\{\delta, \delta_\bp\}=\partial_\bp$ only modulo equations of motion. 
  
   The vector symmetry is a symmetry of  $I_V=I+I'$, but not of I and I' separately. Notice that this   symmetry conserves the  $U(1)$ shadow symmetry only modulo shadow number 2.
   Thus  the vector symmetry connect the $Q$-exact term and the term in the cohomology of $Q$.  In fact, its requirement  forces the above choice of the gauge fermion. For TQFT observables, it can  be relaxed, since the mean values of  $Q$-invariant observables donnot depend on the chosen coefficients in the $Q$-exact terms.
   
   This shows that  the  N=1, d=10 action is determined by a supersymmetry with only 6 =(1 (scalar)  +5  (vector))  parameters\footnote{ This is in agreement with the  former result that   the   N=4, d=4 theory,  that is, the compactification of the N=1,d=10  theory in d=4,  is also   determined by 6  =1 scalar +1 scalar+4 vector  parameters \cite{n=4babo}.}
   \footnote{  
  In  superstring theory \cite{nathan}, there is  also   a six dimensional subalgebra of maximal supersymmetry, with manifest $U(5)\subset SO(10)$  invariance, which   points out  furthermore the relevance of pure spinors.}. The 10 other symmetries occur as ``accidental" extra supersymmetries,  which  enables the untwisting toward the Poincar\'e supersymmetric theory. Most of the proofs concerning the theory should be doable by using the  core-supersymmetry with 6 generators, provided that the 10 other ones have no anomaly, which can be shown to be the case. A superspace with 6 fermionic directions can be clearly constructed, in the line of  \cite{bana}\cite{10DSYM}\cite{ twistsp}.
  
  \section{Topological observables}

 One has a  set of  $Q$-invariant observables because  the cohomology of $Q$ is non empty.

  They follows from   the existence  of descent equations. They are   obtainable by a simple rewriting of the BV equations  for $ \int   \Omega _{5,0}\T (\A  d\A  +\frac{2}{3} \A\A\A)$, as follows  \def\d{\overline \partial} (where $\d\equiv dz^\bm \partial_\bm$)
 \bea
(Q+\d)\A  +  \A\A=0\ \ \ \rm { that\ is } \ \ \ \ \  Q\A = -\d\A  -  \A\A
\eea
 One has 
\bea
Q \T  (\A  \d\A  +\frac{2}{3} \A\A\A)&=&-\T (\d\A  +  \A\A) Q\A= \T (\d\A  +  \A\A)(d \A  +  \A\A)\CR&=&-\d\T (\A  \d\A  +\frac{2}{3} \A\A\A)
\eea
  One has thus 
  \bea
  (Q+\d)\T  (\A  \d\A  +\frac{2}{3} \A\A\A)=0
  \eea
  so that one gets   descent equations  (they are actually very easy to check directly, using the Bianchi identity),
  \bea
Q\T(\chi_{0,2}  \D \chi_{0,2} ) &=&  \d \T( \chi_{0,2} F_{0,2})\CR
Q\T  (\chi_{0,2}  F_{0,2})  &=&    \d\T  (  A_{0,2}  \d A_{0,2}   +\frac{2}{3} A_{0,2} A_{0,2} A_{0,2} )  \CR
Q\T(   A_{0,2}  \d A_{0,2}   +\frac{2}{3} A_{0,2} A_{0,2} A_{0,2}) &=& 0
\eea
 We have therefore the following observables, defined as elements of the cohomology of the scalar supersymmetry
  \bea
  &\int  _{ {\cal{ M}}_{0,5}  } \T  ( \chi_{0,2}  \D \chi_{0,2}  )\CR
&\int _{ {\cal{ M}}_{0,4}  }  \T (  \chi_{0,2}  F_{0,2}) \CR
&  \int  _{ {\cal{ M}}_{0,3}  } \T (   A_{0,2}  \d A_{0,2}   +\frac{2}{3} A_{0,2} A_{0,2} A_{0,2}) 
  \eea
  
%
  All gauge invariant functionals of $A_{0,1}$ are also in the  cohomology of $Q$  since $QA_{0,1}=0$. One has for instance the Wilson loops of the following type
  \bea exp \int dz^\bm A_\bm\eea  
 Because $\Psi _{m}$ and $A_m$ build a trivial  $Q$-doublet, observables cannot depend on  them.

 \section{More on the gauge degeneracy  of $\Tr \chi_{02}\D \chi_{02}$}
 
 In the abelian case, the      $Q$-exact term can be    understood as a gauge-fixing of the  degenerate gauge symmetry  of the action  $\T \epsilon_{mnpqr}
\chi_{\bm\bn} d_\bp \chi_{\bq\br} $, 
 $\chi_{0,2}\sim \chi_{0,2} +\d\epsilon _{01} ,  \epsilon _{01} \sim  \epsilon _{01} +\d \epsilon $  of   the  term  $\T \epsilon_{mnpqr}
\chi_{\bm\bn} d_\bp \chi_{\bq\br} $.  This symmetry leaves    invariant the    
BV fermion ${\cal Z} _V$.
 \def\dd{{\partial}}
We can build a BRST symmetry with  ghosts, antighosts and Lagrange multipliers for $\chi_{02}$, that can be called equivariant with respect to  ordinary $U(1)$ (Yang--Mills) gauge transformations. It needs
 a ghost associated to $\epsilon _{01}$,   that we identify with    $A _{01}$ ($A_\bm$  has the opposite even statistics to $\epsilon _{\bm}$).  We identify  $A _{01}$ and    $\Psi _{10}$ as its antighost and Lagrange multiplier, respectively. 
 
 We thus have the following  equivariant BRST symmetry of  $\chi_{02}\d \chi_{02}$, defined modulo ordinary gauge transformations $A _{01}\sim A _{01}+ \d c$ and  $ A _{10}\sim A _{10}  + \dd c$,
 \bea
 s \chi_{02} &=&\d A _{01}    \ \ \ \ \ \ \ \  \ \ \           s \chi  =   h                      \CR
  s A _{01} &=&0                   \ \ \ \ \ \ \ \    \ \ \ \ \ \ \ \    s h =  0    \CR
   s A _{10} &=&\Psi_{10}       \ \ \ \ \ \ \ \ \ \ \     s \Psi  _{10} = 0  
  \eea
  This is  an abelian BRST symmetry for a (anticommuting) 2-form and one 
  can  thus   build the following  BRST-exact gauge-fixing for the 2-form gauge symmetry, 
  \bea
  s\big (A_{[m} \partial_{n]} \chi_{\bm\bn} +\chi (\frac{h}{2} +\partial _{[\bm} A_{m]})\big)\CR
  =F_{mn} F _{\bm  \bn} 
  + \frac{b^2}{2} + b F_{\bm m}  
  -\chi_{\bm\bn}   \partial_ {[m} \Psi_{n]}-\chi \partial _{[\bm}  \Psi_{m]}
    \eea
 It     is equivariant with respect to the abelian Yang--Mills symmetry. The complete action 
  \bea
  \chi_{02}\d \chi_{02}+ s(A_{[m} \partial_{n]} \chi_{\bm\bn} +\chi (\frac{h}{2} +\partial _{[\bm} A_{m]}))
  \eea
 reproduces therefore  the abelian  version of supersymmetric action (\ref{twisted}) that we  previously  built from other considerations,  using the  BV Chern--Simons action.

  The non-abelian case needs more  refinement.   We only indicate that 
%
it  needs the introduction of a 3-form gauge field, since the  variation of  $\int \Tr  \chi_{02}\D \chi_{02} $ under  
  $ s \chi_{02} =\D \epsilon _{01}$  is 
  $\int \Tr  [ F_{02},  \epsilon _{01}] $, which implies the introduction of a compensating term, resulting into   
  the following classical action  
  \bea
  I_{classical}( \chi_{02}, A_{01},B_{03})=
 \int  \Tr   ( \chi_{02}\D \chi_{02}+   B_{03}F_{02})
 \eea
 Its gauge symmetry involves two non abelian parameters $\epsilon _{01}$ and $\epsilon _{02}$, 
  \bea
 \delta \chi_{02} &=&\D \epsilon _{01}\CR
 \delta B _{03} &=&\D \epsilon _{02} - [ \epsilon _{01}, F_{02}]   
 \eea
 Here $ B _{03}$ must interpreted as a field, not an antifield.  This action can be  a priori quantized in two different ways.  One possibility is a BRST invariant gauge-fixing, which results in a different action than the non-abelian  action~(\ref{twisted}). The other one  is by using  the equation of motion    $F_{02}=0$ of $B _{03}$ and   eliminating $A_{01}$ by solving this equation. By substitution in $\chi_{02}\D \chi_{02}$, this gives   a sort of non-linear  sigma model coupled to $\chi _{02}$.  We will not discuss it here.

    \section*{Acknowledgments}
 I thank  
 G. Bossard for  very interesting and constructive   discussions on the subject and a critical reading of the manuscript. After completion of this work,  N. Nekrasov showed me an unpublished draft \cite{Park}   where many of the
aspects of this publication have been independently perceived.  I also thank N. Berkovits   for useful remarks and insight.

\end{document}